# Can epidemic models describe the diffusion of topics across disciplines?

**Istvan Z. Kiss[1], Mark Broom[1], Paul Craze[2] & Ismael Rafols[3,4*]**

[1] *Department of Mathematics, University of Sussex, Brighton. BN1 9RF, UK*
[2] *Department of Biology and Environmental Science, University of Sussex, Brighton. BN1 9QG, UK*
[3] *SPRU (Science and Technology Policy Research), University of Sussex, Brighton. BN1 9QE, UK*
[4] *Technology Policy & Assessment Center, School of Public Policy, Georgia Institute of Technology. Atlanta, GA 30332, USA.*
[*] *Corresponding author (i.rafols@sussex.ac.uk)*

20th May 2009

**Abstract**

This paper introduces a new approach to describe the spread of research topics across disciplines using epidemic models. The approach is based on applying individual-based models from mathematical epidemiology to the diffusion of a research topic over a contact network that represents knowledge flows over the map of science –as obtained from citations between ISI Subject Categories. Using research publications on the protein class *kinesin* as a case study, we report a better fit between model and empirical data when using the citation-based contact network. Incubation periods on the order of 4 to 15.5 years support the view that, whilst research topics may grow very quickly, they face difficulties to overcome disciplinary boundaries.

**1. Introduction**

How concepts, ideas, technologies and/or innovations spread across heterogeneous communities has long been one of the central questions of the sociology of science and technology (Rogers, 1962; Mulkay, 1974). In recent years, studying the diffusion of scientific topics has become much more feasible due to the wider availability of a variety of databases, fast and cheap computing power and efficient search and model-fitting algorithms. There are a number of ways in which the diffusion of topics can be tracked (e.g. Chen and Hicks, 2004). In terms of network dynamics, the similarities between the spread of research topics and the spread of infections diseases have not gone unnoticed (Bettencourt *et al.*, 2006). In the spread of a disease through a population, contact between an infectious and a susceptible individual can lead to the transmission of infection. In a similar way, individuals or groups working on a particular research topic or topics can motivate other individuals or groups to start work based on the same or similar research topics with citation being evidence of motivation.

Though models of social contagion date back to the mid 20$^{th}$ century[1], the use epidemiological models for to model diffusion in scientific publications was recently discussed by Bettencourt *et al.* (2006), who found a good fit between suitably adapted epidemic models and data for the spread of a specific research topic (Feynman diagrams in theoretical physics). They further showed that this good fit is not dependent on the particular topic chosen and that epidemic models provide good descriptions of the spread of other topics in both theoretical and experimental physics (Bettencourt *et al.* 2008).

However, the epidemiological models investigated so far have been of the simple differential-equation-based compartmental type. While compartmental models are transparent and allow the derivation of some analytical results, they are limited in their capability to capture heterogeneities at the individual level and in the interaction between individual epidemiological units, both of which we expect to see in citation networks (see model description below). As a potentially useful alternative method, we have developed an individual-based weighted network model.

The second novelty of the approach we present here is that, whereas previous studies have investigated the growth of a topic in terms of number of published papers or publishing authors, we inquire here into how a research topic spreads over an existing network of disciplines. In other words, whereas previous studies had focused on growth dynamics, this novel perspective captures the diffusion of topics over network of connections between scientific disciplines, as assigned by the ISI Web of Science's classification in terms of Subject Categories (SCs), following Leydesdorff and Rafols' approach (2009). This underlying network of citations among SCs represents the knowledge flows over the "backbone" of the map of science (Boyack et al., 2005). The weight of a link (i.e., the normalised number of citations between SCs) in this network is taken to be a good indicator of the likelihood of a SC becoming research-active in a certain area given that some other related SCs are already research-active in this specific area. We can then ask whether a novel topic (a newly discovered phenomenon, material, or method or piece of instrumentation) seeded at one particular point in the network will diffuse through it following to some extent the weighted connections between SCs.

In this exploratory study we examined the spread of research on *kinesin* (often referred to as a molecular motor or "nano-engine"). Kinesin represents a class of eukaryotic motor protein that functions by moving actively along microtubules (Block, 1998). Kinesin research first emerged in 1985, with the report of its discovery published in the areas of Biochemistry and Cell Biology. In the 1990s research on kinesin spread broadly to other fields in the biological sciences and in the 2000s it reached other biomedical research, on the one hand, and chemistry, physics and materials sciences, on the other, as illustrated in Figure 1. This later development is associated with potential bionanotechnology applications, what made kinesin an interesting case for the study of interdisciplinary research (Rafols, 2007, Rafols and Meyer, 2009).

---
[1] See historical review in the introduction of Bettencourt et al. (2006).

Here, we show that the spread of kinesin-related research over a network of disciplines can be well approximated by models used in the context of the transmission of infectious diseases (Anderson & May, 1991; Diekman & Heesterbeek, 2000; Keeling & Rohani, 2008). Similar network models have been successfully used to explain and predict the pattern of infectious disease transmission (Keeling et al., 2001; Green et al. 2008; Kiss et al. 2005, 2006a, 2006b), and such models are well researched in the context of mathematical epidemiology (Keeling & Eames, 2005).

The paper is organised as follows: we first introduce the data and methods; second, we describe the model; in section 4 we present the results of the quality of fit for two different disease transmission models (i.e., Susceptible-Infected or *SI,* and Susceptible-Exposed-Infected or *SEI*). Results based on the weighted empirical network are compared to the case of homogeneous disciplinary spread (i.e., equal weights) on the same network. The discussion and conclusions briefly explore possible future improvements of the model and its applications in science policy.

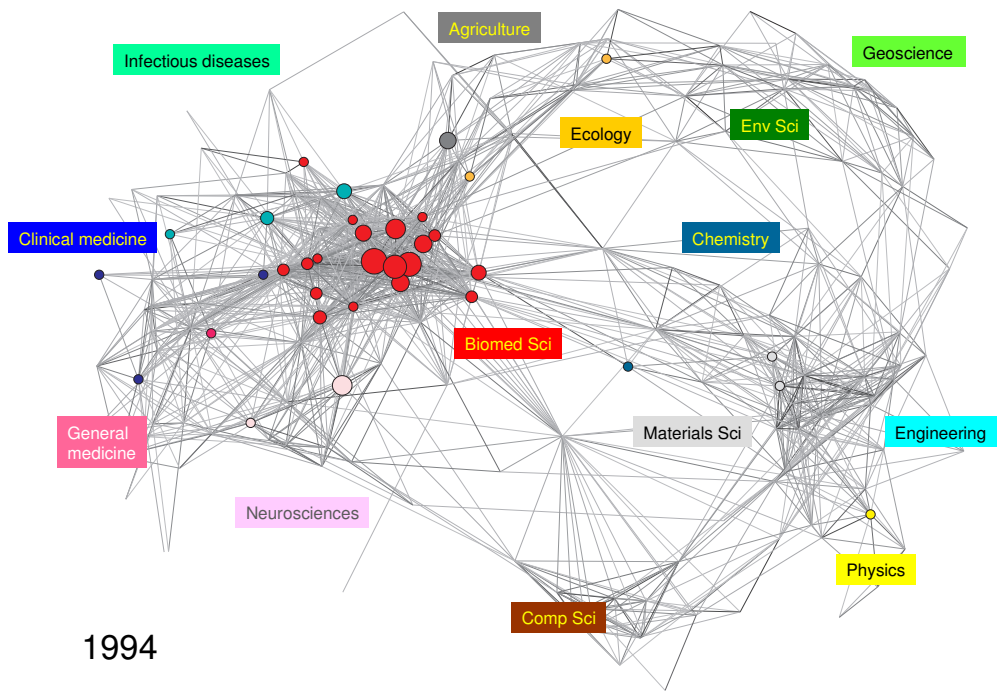

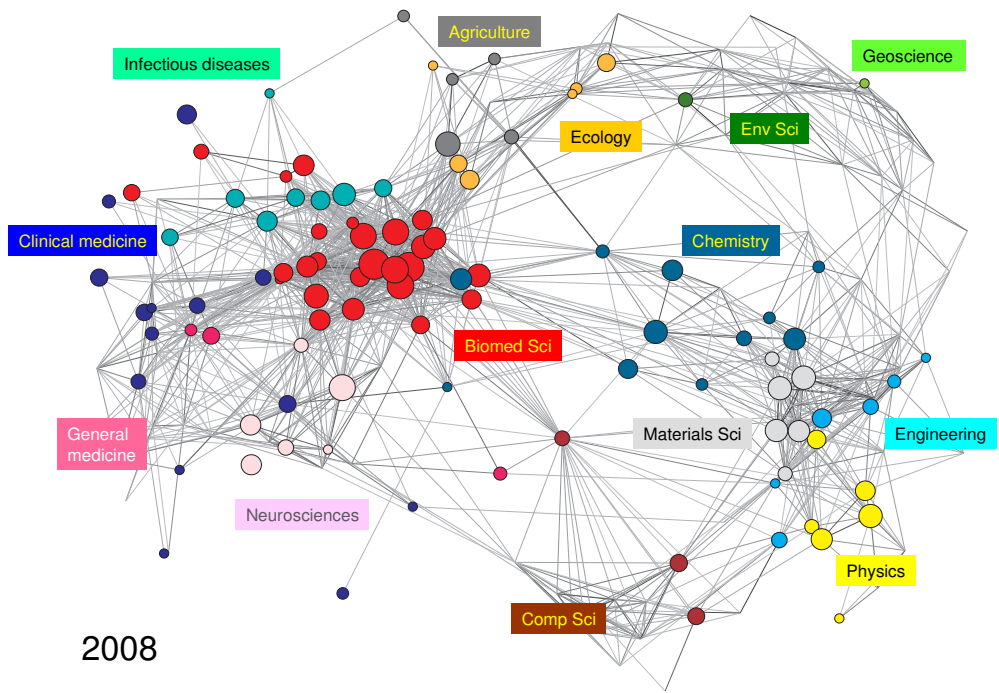

Figure 1: Cumulative number of publications related to kinesin in the map of science for 1994 and 2008. Each node represents a Subject Category (SC). The size of nodes is proportional to the number of publications. The positions of SCs are related to similarities in citations as described in Leydesdorff and Rafols (2009). The weighted network of links aims to illustrate the citation flows among SCs.

## 2. Methods and data

A set of publications (articles, reviews and letters) related to the molecular motor kinesin was constructed searching the term "kinesin" in the bibliographic field "Topic" of the ISI Web of Science database. This search yielded 4,021 publications starting from 1985 (2 publications) to 2007 (394)[2]. Each publication was assigned to one or more disciplines according to ISI Web of Science's classification in terms of SCs. The matrix of cross-citations between SCs was obtained from Leydesdorff and Rafols' (2009)[3]. This SC to SC citation matrix had been created for 2006 from the Journal Citation Reports (JRC) of the Science Citation Index (SCI). This matrix of cross-citation among disciplines is very dependent on how the disciplines (here SCs) are defined, an issue on which there is little agreement. However, comparisons of science maps by Klavans and Boyack (2009) and Rafols and Leydesforff (2009) suggest that even for very different classifications the basic characteristics of the overall structure of the science matrix are quite robust[4].

This citations matrix of 171 SCs as nodes ($N=171$)[5] was used to construct the contact network over which the transmission dynamics unfolds (in this case, the spread of research topics). This baseline citation network may be understood as representing the knowledge flows among SCs. For example, a citation from $SC_j$ to $SC_i$ represents a potential route on which knowledge could have spread from $SC_i$ to $SC_j$. The key assumption in the model is that that the weight of a link of this network is a good indicator in determining the likelihood of a SC becoming research-active in a certain area given that some other related SCs are already research-active in this area. The links were normalised so that the weight of the incoming links for all SCs add up to one. Hence the weight $w_{ij}$ of the directed link from $SC_i$ to $SC_j$ is given by:

$$w_{ij} = \frac{\text{\# of citations from } SC_j \text{ to } SC_i}{\text{\# of total citations made by } SC_j} \quad (1)$$

with $\sum_i w_{ij} = 1$ for $\forall j = 1 \ldots N$ (2).

The distribution of link weights is skewed and close to a scale-free distribution (Barabási & Albert, 1999). The directed and weighted SC network was found to be highly connected with many links but most of them with very small weights. Further descriptive details of the network are presented in the supplementary information.

---

[2] Due to improved indexing since 1991, this search underestimates the number of publications until 1990 – an effect we will overlook here.
[3] The matrix of SC-to-SC cross-ctiations is available at Loet Leydesdorff's webpage: http://www.leydesdorff.net/map06/data.xls .
[4] In spite of the use disparate classifications and methods to create and portray the maps of science, their overall structure generally bear striking similarities (Boyack et al. 2005; Moya-Anegón et al., 2007; Rosvall and Bergstrom, 2008; Klavans and Boyack, 2009; Rafols and Leydesdorff, 2009).
[5] The SCI had 172 SCs in 2006. We removed the SC "Multidisciplinary Sciences" because we understood that it might lead to misleading linkages, given that the publication patterns of journals such as *Nature* or *Science* publish for diverse audiences but do not necessarily *connect* them.

## 3. The model

In classic disease transmission models with the assumption of homogeneous random mixing, the population is divided into different compartments based on the disease status of the individuals and other characteristics such as age, gender or risk. Thereafter, the rates of all possible transitions between the compartments are determined. Based on this, a system of differential equation can be derived. In our model, we use a different approach and consider each SC as a node in a network along with all its weighted connections to other nodes or SCs. Based on Sharkey (2008) and Kiss et al. (2005, 2006a, 2006b) we use an individual-based model where equations for the probability of being in a particular state (e.g. susceptible, *S;* exposed or latent and incubating, *E;* or infected and infectious, *I*) at a particular time are worked out based on the links between SCs, the status of neighbours, and given transmission and transition rates. SCs that are susceptible (S) are either not aware of a particular research topic or, if aware, may still not adopt it. Incubating SCs (*E*) are those that are aware of a certain topic and have moved onto actively engaging with it. This is expected to result in tangible research output in the form of papers. Infected SCs or adopters (*I*) are those that are actively working and publishing in a particular research topic. Further states such as recovered (i.e., SCs that have stopped working on a particular research area, often denoted by *R*) and sceptics or stiflers (i.e., SCs that are aware of the topic but do not adopt it or prefer another competing topic, often denoted by *Z*) are possible. In our current model the recovered and sceptics states are not considered.

We examined two models, a Susceptible-Exposed-Infected (*SEI*) model and a simpler Susceptible-Infected (SI) model. The *SEI* model equations are given by:

$$\begin{cases} dP_{S(i)}(t)/dt = -\sum_j T_{ji} P_{I(j)}(t) P_{S(i)}(t), \\ dP_{E(i)}(t)/dt = \sum_j T_{ji} P_{I(j)}(t) P_{S(i)}(t) - g P_{E(i)}(t), \quad (3) \\ dP_{I(i)}(t)/dt = g P_{E(i)}(t), \end{cases}$$

where $0 \leq P_{I(j)}(t) \leq 1$ denotes the probability of node *j* being infected at time *t* (likewise for *E(j)* and *S(j)*). Throughout the simulation, $P_{S(j)}(t)+P_{E(j)}(t)+P_{I(j)}(t)=1$, for all $\forall t>0$. The contact network is represented by $T_{ji} = \tau G_{ji}$ with $G_{ji}=(w_{ji})_{j,i=1,...,N}$ denoting the adjacency matrix that includes link weights. $\tau$ is the transmission rate per contact and $1/g$ is the average incubation or latent period. By numerically integrating the ordinary differential equations, the number of the infected or adopter SCs at time *t*, according to the model, can be estimated as $I(t) = \sum_j P_{I(j)}(t)$.

The *SEI* model (Equation 3) can be simplified to the case of an *SI* model where the possibility of an exposed period is excluded. The equations for the simple *SI* model are

$$\begin{cases} dP_{S(i)}(t)/dt = -\sum_{j} T_{ji} P_{I(j)}(t) P_{S(i)}(t), \\ dP_{I(i)}(t)/dt = \sum_{j} T_{ji} P_{I(j)}(t) P_{S(i)}(t). \end{cases} \quad (4)$$

The simulations for both models were started at time $t = 0$ corresponding to 1985 and the equations were integrated forward in time until 2007. The initial infection was seeded in the two SCs corresponding to *Biochemistry and Molecular Biology* and *Cell Biology*. The *SEI* model has two free parameters that allow fitting of the model output to the empirical data:

- $\tau$, the per contact transmission rate.
- $g$, where $1/g$ is the average incubation or latent period.

In the *SI* model, only $\tau$ was estimated. In both cases the cumulative SC count and $I(t)$ were normalised by $N$ and compared on a yearly basis. The estimation of parameters was performed based on a modified version of the Kolmogorov-Smirnov statistic. That is, a minimum distance estimation between an empirical distribution function of a sample and the cumulative distribution function of the reference distribution:

$$AdaptedKS = \sup_{t=1985,1986,\ldots,2007} \left[ \frac{1}{N} \left( \sum_{j=1}^{N} P_{I(j)}(t) - EmpiricalCount(t) \right) \right]. \quad (5)$$

The count (cumulative) of the SCs that had become active in kinesin-related research provides information at the level of all SCs or population level. We note that although this is good method to estimate the parameters, we cannot assess the fit by performing the actual Kolgomorov-Smirnov test, as the contributing data are not independent and identically distributed (i.i.d.). Apart from accurate prediction of the growth in the number of SCs, an appropriate model that fits the data well can also predict the exact SCs that are active, at a particular time, in kinesin-related research. To monitor model prediction at the SC level the following likelihood function is applied:

$$L = \frac{1}{M} \sum_{t=1985,1986,\ldots,2007} \left( -\log \left( \prod_{i=1}^{N} \left( P_{I(i)}(t) \right)^{Y_i(t)} \left( 1 - P_{I(i)}(t) \right)^{1-Y_i(t)} \right) \right), \quad (6)$$

where $Y_i(t) \in \{0,1\}$, is an indicator function with a value of one denoting a SC that was active in kinesin-related research at time $t$ and zero if otherwise (Keeling et al. 2001, Green et al. 2008). $M$ denotes the number of time points where comparisons at the individual level were made. In this case, $M = 23$ and this corresponds to yearly comparisons from 1985 until 2007. Similarly, to above, we cannot perform the standard likelihood ratio test to asses the model fit, as data are not i.i.d.

**4. Results**

This section presents the results of the simulations. First we examined the SI model with the empirical weighted network. Second, we explored how results are affected by stronger simplifications on the base network. Finally, we investigated the *SEI* model.

**4.1. Susceptible-Infected model**

*Weighted network*

For this simple one parameter model, the fit to the empirical data was good and both measures of model fit were minimised for the same value of $\tau$ (figure 2), correct to three decimal places. This indicates that this simple baseline model captured the spread of kinesin-related research to a good degree. The model output slightly overestimated the initial growth but did better for later years.

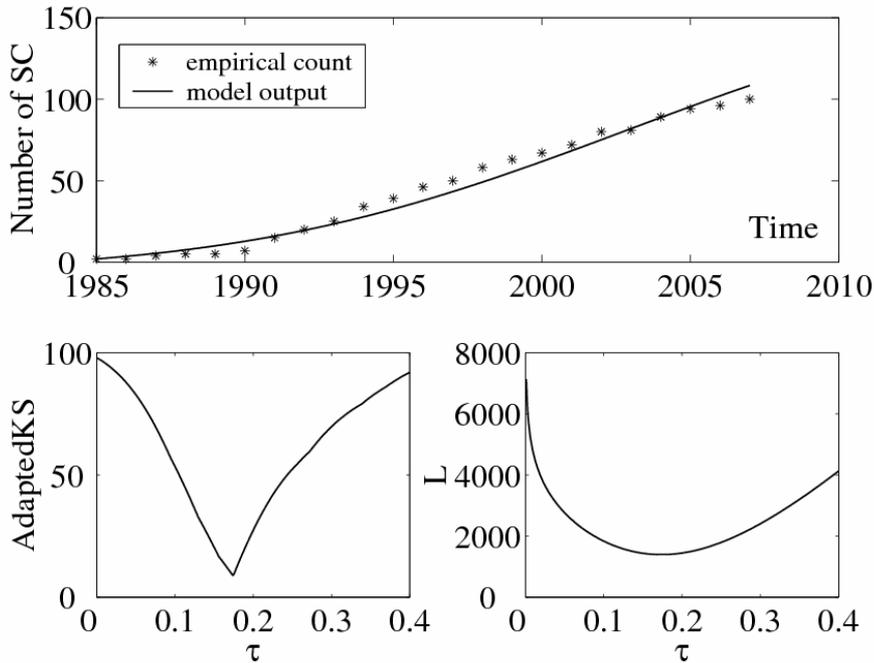

Figure 2: Best fit curves of the growth of the number of SCs that are active in Kinesin-related research (top panel) and *AdaptedKS* (bottom left) and *L* (bottom right) as a function of $\tau$. Model based on the weighted network and *SI* type transmission. Best fit for both measures is obtained for $\tau = 0.174$ with the corresponding *AdaptedKS*=0.051885 (8.872447 in terms of counts) and *L*=1395.716852. The optimal value of $\tau$ is given up to a $10^{-3}$ precision.

*Exploring variations in the weights*

To explore the significance of the empirical weight distribution in explaining the spread of research topics, two other link weight distributions were considered. First the case where all links were equal to the average link weight; second, the case when the weights of all incoming links of any node or SC are equal and sum to one.

*i. All weights equal*
First, we considered the weighted network case where all link weights are assumed to be equal to the average link weight over the whole network ( $w_{ij} = 171/22295$ for $\forall i, j = 1,\ldots,N$ ). In figure 3, the best fit case is illustrated with an optimal value of $\tau$ that is comparable to that obtained from the weighted network case. While the initial fit up to 1990 is very accurate, for later years the fit is less accurate when compared to the weighted network model. This indicates that weights based on the citation pattern are important in understanding and modelling the dynamics of the spread. The importance of weights is further emphasised by the higher values of *AdaptedKS* and *L* compared to the weighted network case.

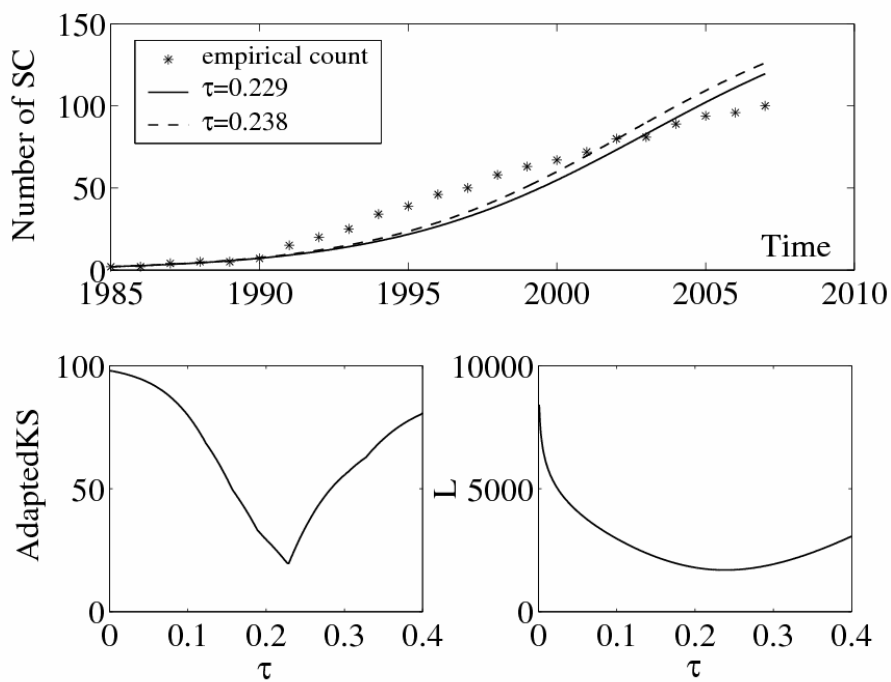

Figure 3: Best fit curves to the growth of the number of SCs that are active in Kinesin-related research (top panel), and *AdaptedKS* (bottom left) and *L* (bottom right) as a function of $\tau$. Model based on a network with all weights equal to the average weight across all links of the original network and with an *SI* type transmission. $\tau = 0.229$ minimises *AdaptedKS* (=0.114695 or 19.612834 in terms of counts) with a corresponding value of *L*=1700.530081. $\tau = 0.238$ minimises *L* (=1694.533715) with a corresponding value of *AdaptedKS*=0.153308 (26.215715 in terms of counts). The optimal values of $\tau$ are given up to a $10^{-3}$ precision.

*ii. Weights of all incoming links of a node equal and sum to one*
In this second case the network is weighted such that the weights of links pointing towards any node are all equal to the inverse of the destination node's in degree. (e.g., the weight of all links that point to a node with an in degree equal to 10 will be equal to 1/10=0.1). The results of this case are very similar to the previous one. As shown in Table 1, the latter weight distribution performs marginally better in minimising the value of *AdaptedKS* but does clearly better in minimising *L*.

In conclusion, the simple SI model shows quite good agreement with empirical data for kinesin, but this agreement can be shown to depend on the use of the weighted network. This result supports the main idea behind this paper: that diffusion of topics on the map of science is more likely to occur between disciplines with existing knowledge flows.

**4.2 Susceptible-Exposed-Infected model**

An important component of the transmission of topics is the latent or incubation period $(1/g)$ that represents the time needed to assimilate and apply research topics. The *SEI* model has two parameters, with the latent period having an important effect on the initial growth rate of the number of SCs becoming active, $r$ (i.e., $I(t) = ce^{rt}$). In figure 4 we illustrate the best fit prevalence curves based on the *AdaptedKS* and *L*. To interpret these results, it is useful to think in terms of first considering a fixed latent period $(1/g)$ and thereafter determining the value of $\tau$ that minimises the difference between data and model output. Long latency periods delay the infection and many infected individuals remain exposed for longer. Thus, to get a reasonable fit, high values of the transmission rate $\tau$ are required. This tendency is reflected by a set of optimal parameter pairs $(1/g, \tau)$ with both latent period and transmission rate increasing simultaneously (figure 4, bottom left and right). However, the quality of fit, along the set of optimal pairs, changes with the best agreement based on *AdaptedKS* occurring for $(1/g, \tau) = (15.5, 1.90)$. For longer latent periods this measure indicates that the discrepancy between model output and data increases (figure 4, bottom left). A similar tendency is observed when the parameter estimation happens based on *L* with the best agreement between data and model output for $(1/g, \tau) = (4.0, 0.37)$.

The minimum value of the *AdaptedKS* is considerably lower than that corresponding to the *SI* model. The minimum value of *L* is also clearly lower, although this is obtained for a parameter pair that is very different compared to the pair that minimised *AdaptedKS*. While the agreement at the population level is much better for the *SEI* model, for the same pair of parameters, the agreement at the individual level is not as good as for the simple *SI* model. The same observation is valid when considering the minimum value of *L*. Hence, in the two parameter model agreement at both individual and population level is difficult to obtain. This difficulty is indicative that the model may be improved further as described in the next section.

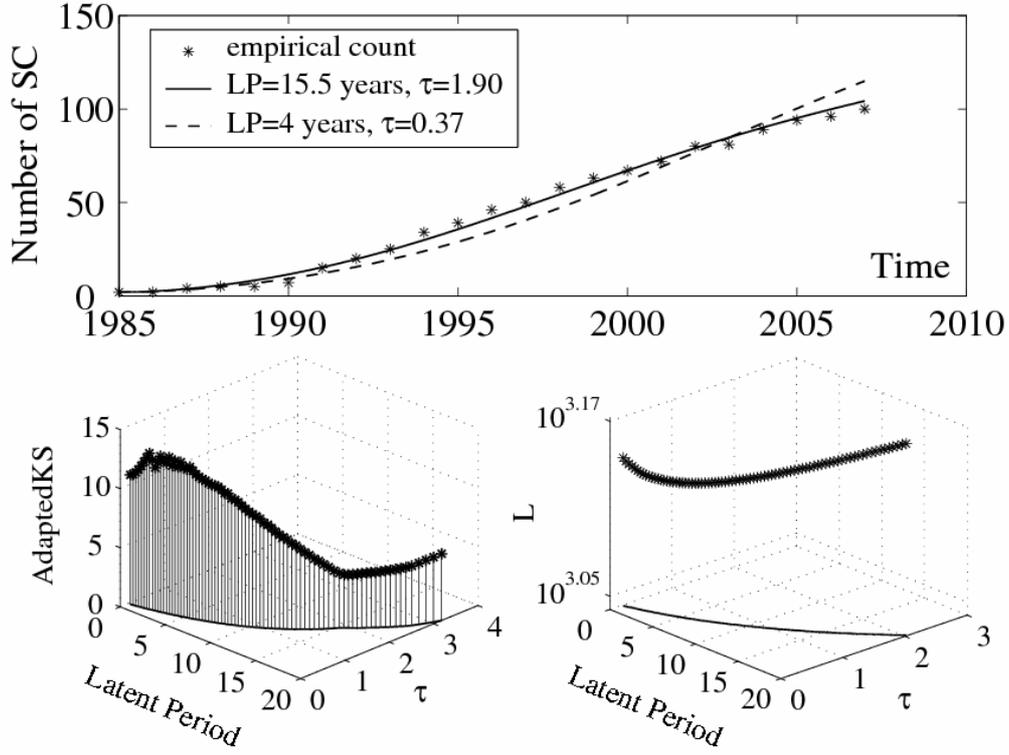

Figure 4. Best fit curves to the growth of the number of SCs that are active in Kinesin-related research (top panel), and *AdaptedKS* (bottom left) and *L* (bottom right) as a function of the latent period and $\tau$. Model based on the weighted network with *SEI* type transmission. A latent period of 15.5 years and $\tau = 1.90$ minimises *AdaptedKS* (=0.026186 or 4.477749 in terms of counts) with a corresponding value of *L*=1460.664702. A latent period of 4 years and $\tau = 0.37$ minimises *L* (=1358.911193) with a corresponding value of *AdaptedKS*=0.087254 (14.920351 in terms of counts). The optimal values of $\tau$ are given up to a $10^{-2}$ precision with the latent period to a precision of ¼ years.

| **Model** | **Network Type** | $\tau$ | $g$ | *AdaptedKS* | *L* |
|---|---|---|---|---|---|
| *SI* | Empirical weighted | 0.174 | NA | Min=0.051885 | Min=1395.716852 |
| *SI* | All weights equal | 0.229 | NA | Min=0.114695 | 1700.530081 |
|  |  | 0.238 | NA | 0.153308 | Min=1694.533715 |
| *SI* | Weights of all incoming links sum to one | 0.230 | NA | Min=0.115324 | 1811.015665 |
|  |  | 0.238 | NA | 0.148455 | Min=1806.257798 |
| *SEI* | Empirical weighted | 1.90 | 1/15.5 | **Min=0.026186** | 1460.664702 |
|  |  | 0.37 | 1/4 | 0.087254 | **Min=1358.911193** |

Table 1: Summary of parameter estimates for different network and disease transmission models. In bold we indicate the best fit model and optimal parameters.

## 5. Discussion

The results of the simulations showed that relatively simple models with the weighted network of SCs can produce good fits and deliver estimates of transmission rates and incubation times. As the present study is essentially a proof of concept, these results should be treated with a degree of caution but they do suggest directions in which this methodology could be further developed. We envisage two aspects in which the models might be incrementally improved.

First, we have deliberately restricted this initial study to the two simplest epidemiological models, with empirical data that only considers whether SCs have been active in kinesin research. While the transmission among SCs is crucial in the initial 'seeding' stages, after a SC started work on a particular research topic, the main driver of growth may come from activity within that particular SC. Hence, some form of within SC dynamics that goes beyond dichotomous description of SCs and takes into account its degree of activity (i.e. its relative amount of publications it has in the topic) can be important if trying to improve the model fit.

Second, whilst here we have used a cumulative description, assuming that one active/infected SC never lost its activity in the field, changing the type and/or number of compartments may result in a better description of the observed spread. For example, allowing SCs to "forget" kinesin after some period without publications seems a realistic assumption. This would echo the findings of Bettencourt *et al.* (2008), where elaboration of their simple epidemiological model to include recovered (i.e. researchers who have produced papers on a particular topic but have then moved on to other research) and sceptic (i.e. researchers who stifle or prevent the movement of ideas they do not accept) classes resulted in better descriptions on their empirical data.

Nevertheless, the good quality fits obtained in the simulations also suggest that even the simple models may already provide insights on the dynamics of science. From this perspective, the values of transmission and incubation rates obtained indicate that diffusion over disciplines takes a considerable time: in the range of 0.53 to 2.70 years for a transmission per contact (i.e., $\tau$ between 0.37 and 1.90) and 4.0 to 15.5 years for the incubation period. These results would support the view, in agreement with many qualitative findings, that the crossing of disciplinary boundaries takes considerable time. On this direction, the obvious extension of the current study is to compare transmission and incubation rates between different topics or emergent fields, in particular for areas as bionanotechnology that are construed as highly interdisciplinary (Rafols, 2007). These studies may then be used to test the claims of radical changes in the dynamics of science (Etzkowitz and Leydesdorff, 2000; Bonaccorsi, 2008). As an alternative, on more practical grounds, the analyses can be useful to inform policy makers whether (and how) theoretical methods (e.g., Feynmann diagrams) spread more quickly than those requiring a large investment of experimental equipment (e.g., nanofabrication) even when the underlying social and cognitive networks are quite similar.

## 6 Conclusions

This paper has demonstrated the feasibility of applying individual-based epidemic models to the spread of a research topic over the map of science. It has made two contributions beyond previous epidemic models (Betterncourt et al., 2006, 2008): the use of a weighted network of disciplines to describe the spread of topics, and the introduction of individual-based models. Using research on kinesin as a case study, we have confirmed that the agreement between model output and empirical data significantly increases when the normalised weighted contact network between SCs is used (the base map of science). The investigation has allowed us to discuss possible further improvement in the models, e.g. by considering internal SC growth dynamics (e.g. taking into account not only whether a SC is infected, but also how active it is) or loss of activity (recovery) of a SC.

Although this is a proof of concept study and results need to be treated with caution, the incubation periods obtained, on the order of 4 to 15.5 years, support the view that, whilst research topics may grow very quickly, they face difficulties to overcome disciplinary boundaries. This type of information regarding the diffusion rate of research topics over disciplines may be of particular interest for emergent fields such as nanotechnologies to test claims (and hype) of radical changes in knowledge dynamics (Bonaccorsi, 2008).

**Acknowledgements**

IR thanks Loet Leydesdorff for fruitful discussions. IR acknowledges support from the US National Science Foundation (Award #0830207, "Measuring and Tracking Research Knowledge Integration"). The findings and observations contained in this paper are those of the authors and do not necessarily reflect the views of the National Science Foundation.

# Supplementary Information

# Can epidemic models describe the diffusion of topics across disciplines?

**Istvan Z. Kiss[1], Mark Broom[1], Paul Craze[2] & Ismael Rafols[3,4*]**

[1] *Department of Mathematics, University of Sussex, Brighton. BN1 9RF, UK*
[2] *Department of Biology and Environmental Science, University of Sussex, Brighton. BN1 9QG, UK*
[3] *SPRU (Science and Technology Policy Research), University of Sussex, Brighton. BN1 9QE, UK*
[4] *Technology Policy & Assessment Center, School of Public Policy, Georgia Institute of Technology. Atlanta, GA 30332, USA.*
[*] *Corresponding author (i.rafols@sussex.ac.uk)*

## 1. Descriptive statistics of base network

The directed and weighted SC network is a highly connected network with the average number of connections per subject category $\langle k \rangle \cong 130.38$. We define the in and out degree of $SC_i$ as the number of incoming $\left(k_{in}^i\right)$ and outgoing $\left(k_{out}^i\right)$ citations/links respectively. Every directed link has an origin and a destination SC. Hence, the average in degree $\left(\langle k_{in} \rangle = \langle k \rangle = \frac{1}{N}\sum_{i=1}^{N} k_{in}^i\right)$ and the average out degree $\left(\langle k_{out} \rangle = \langle k \rangle = \frac{1}{N}\sum_{i=1}^{N} k_{out}^i\right)$ are equal. The minimum in and out degree in the SC network is equal to 40 and 46 respectively. Both the maximum in and out degree is equal to $N = 171$. The in and out degree distributions (figure S1) show that all subject categories are well connected with a high number of cross-citations. This network also accounts for self-loops denoting citations of papers within a SC by papers published in the same SC. Self-loops are important and represent a significant difference compared to disease transmission models where self-loops cannot spread the infection. In contrast, researchers in a particular SC can motivate or determine other researchers in the same SC to start work based on particular research ideas.

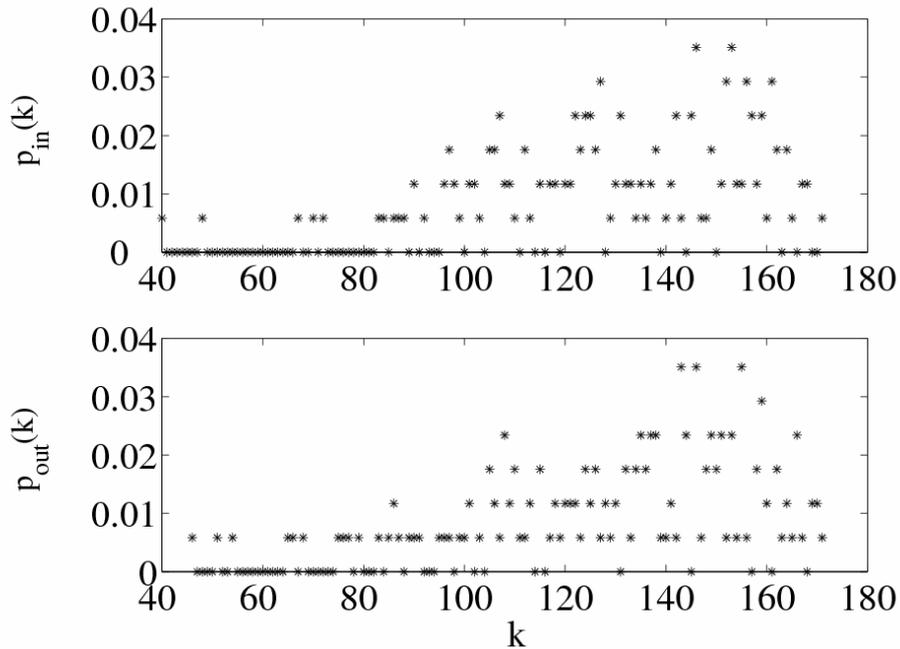

Figure S1: In and out degree distribution of the SC network showing the proportion of SCs with a particular in and out degree count.

The link weights are highly heterogeneous with a close to scale-free distribution (figure S2) (Barabási & Albert, 1999)). This emphasises that while there are a high number of links many are rather weak with very small weights. In many applications weighted networks are significant since it is very unlikely that all links are equally important (Onnela et al., 2007). Many studies assume equal weights in order to allow the derivation of some analytical results or to reduce the complexity and time of simulations. However, there is a clear need to use weighted networks especially when these are used as the backbone for various dynamic processes such as power grid, transportations networks, disease transmission and others. While weights will increase model accuracy, transparency will suffer and analytical results will be difficult to obtain.

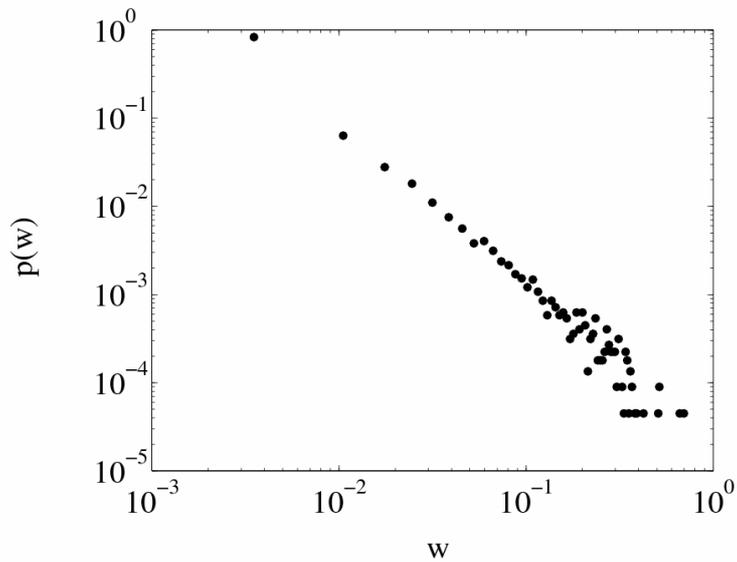

Figure S2: The weight distribution of all links (22295) from the SC network based on citations cumulated over 2006. Distribution based on bins of equal size with bin centres and proportions plotted on a log-log scale.